\documentclass[aps,epsf,nofootinbib,notitlepage,showpacs,10pt, unsortedaddress, superscriptaddress]{revtex4-1}

\usepackage[svgnames, table]{xcolor}

\usepackage{color}
\usepackage{float}
\usepackage{graphicx}
\usepackage{subcaption}
\usepackage{amsmath}
\usepackage{amsthm}
\usepackage{mathtools}

\theoremstyle{definition}

\usepackage{bm}
\usepackage{hyperref}
\usepackage[capitalise]{cleveref}

\usepackage{float}
\usepackage{multirow}
\restylefloat{table}
\usepackage{algorithm}
\usepackage{algpseudocode}

\newcommand{\bra}[1]{{\left\langle #1 \right|}}
\newcommand{\ket}[1]{{\left| #1 \right\rangle}}

\usepackage{quantikz}
\theoremstyle{definition}

\definecolor{AW}{HTML}{286D8C}

\usepackage{tikz} 
\usetikzlibrary{matrix,positioning}
\usepackage{amsfonts}
\usepackage{amssymb}

\algnewcommand{\To}{\textbf{To }}
\algnewcommand\Input{\item[\textbf{Input:}]}%
\algnewcommand\Output{\item[\textbf{Output:}]}%

\definecolor{rpLightPurple}{HTML}{232136}
\definecolor{rpDarkPurple}{HTML}{1F1D2E}
\definecolor{rpWhite}{HTML}{e0def4}
\definecolor{rpPink}{HTML}{c4a7e7}
\definecolor{rpBlue}{HTML}{3e8fb0}
\definecolor{rpLightBlue}{HTML}{9ccfd8}
\definecolor{rpYellow}{HTML}{f6c177}
\definecolor{rpRed}{HTML}{eb6f92}
\definecolor{rpGreen}{HTML}{31748f}
\definecolor{rpBlack}{HTML}{26233a}
\definecolor{rpOtherBlack}{HTML}{6e6a86}
\definecolor{rpMagenta}{HTML}{c4a7e7}

\hypersetup{
   colorlinks = true,
    urlcolor=blue
}
\usepackage[normalem]{ulem}
\usepackage{lipsum}

\graphicspath{{figures/}}
\usepackage{tikz-network}
\usetikzlibrary{fit, shapes.arrows}
\usetikzlibrary{shapes.geometric, arrows}
\usetikzlibrary{positioning,chains,fit,shapes,calc}

\tikzstyle{boxblue} = [rectangle, rounded corners, minimum width=3cm, minimum height=1cm,text centered, draw=black, fill=blue!30]
\tikzstyle{boxred} = [rectangle, rounded corners, minimum width=3cm, minimum height=1cm,text centered, draw=black, fill=red!30]
\tikzstyle{boxgreen} = [rectangle, rounded corners, minimum width=3cm, minimum height=1cm,text centered, draw=black, fill=green!30]
\tikzstyle{io} = [trapezium, trapezium left angle=70, trapezium right angle=110, minimum width=3cm, minimum height=1cm, text centered, draw=black, fill=orange!30]
\tikzstyle{process} = [rectangle, minimum width=3cm, minimum height=1cm, text centered, draw=black, fill=orange!30]
\tikzstyle{decision} = [diamond, minimum width=3cm, minimum height=1cm, text centered, draw=black, fill=green!30]
\tikzstyle{arrow} = [thick,->,>=stealth]

\tikzstyle{decision} = [diamond, draw, fill=blue!20, 
    text width=4.5em, text badly centered, node distance=3cm, inner sep=0pt]
\tikzstyle{block} = [rectangle, draw, fill=blue!20, 
    text width=5em, text centered, rounded corners, minimum height=4em]
\tikzstyle{line} = [draw, -latex']
\tikzstyle{cloud} = [draw, ellipse,fill=red!20, node distance=3cm,
    minimum height=2em]

\usepackage{float}

\usepackage{array, booktabs, makecell, multirow}

\usepackage{amsmath,calc}

\definecolor{graphPurple}{RGB}{185,75,185}

\begin{document}

\markboth{Anthony Wilkie}
{Quantum approximate optimization algorithm with random and subgraph phase operators}

\title{Quantum approximate optimization algorithm with random and subgraph phase operators}

\author{Anthony Wilkie}
\affiliation{
	Department of Industrial and Systems Engineering\\ University of Tennessee at Knoxville\\Knoxville, TN 37996}

\author{Igor Gaidai}
\affiliation{
	Department of Industrial and Systems Engineering\\ University of Tennessee at Knoxville\\Knoxville, TN 37996}

\author{James Ostrowski}
\affiliation{
	Department of Industrial and Systems Engineering\\ University of Tennessee at Knoxville\\Knoxville, TN 37996}

\author{Rebekah Herrman}\thanks{corresponding author}
\email{rherrma2@utk.edu}
\affiliation{
	Department of Industrial and Systems Engineering\\ University of Tennessee at Knoxville\\Knoxville, TN 37996}

\begin{abstract}
    The quantum approximate optimization algorithm (QAOA) is a promising quantum algorithm that can be used to approximately solve combinatorial optimization problems.
    The usual QAOA ansatz consists of an alternating application of the cost and mixer Hamiltonians.
    In this work, we study how using Hamiltonians other than the usual cost Hamiltonian, dubbed custom phase operators, can affect the performance of QAOA.
    We derive an expected value formula for QAOA with custom phase operators at $p = 1$ and show numerically that some of these custom phase operators can achieve higher approximation ratio than the original algorithm implementation.
    Out of all the graphs tested at $p=1$, 0.036\% of the random custom phase operators, 75.9\% of the subgraph custom phase operators, 95.1\% of the triangle-removed custom phase operators, and 93.9\% of the maximal degree edge-removed custom phase operators have a higher approximation ratio than the original QAOA implementation.
   Furthermore, we numerically simulate these phase operators for $p=2$ and $p=3$ levels of QAOA and find that there exist a large number of subgraph, triangle-removed, and maximal degree edge-removed custom phase operators that have a higher approximation ratio than QAOA at the same depth.
    These findings open up the question of whether better phase operators can be designed to further improve the performance of QAOA.
\end{abstract}

\maketitle
\section{Introduction}\label{sec:intro}

The quantum approximate optimization algorithm (QAOA) is an algorithm consisting of a parameterized quantum circuit that is well-suited to approximate solutions to combinatorial optimization (CO) problems \cite{Farhi2014FQAOA}.
The algorithm evolves a given input state under the alternating action of cost and mixing Hamiltonians for variable amounts of time that are selected to optimize the expectation of the cost Hamiltonian.
QAOA can be used to find an approximate solution to any CO problem that can be formulated as an Ising problem \cite{Ozaeta_2022}, among which the most studied is the MaxCut problem \cite{herrman2021impact, farhi2020quantum, FixedAngleWurtz, Guerreschi_2019,wang2018quantum}. The goal of MaxCut is to partition the vertices of a given graph $G=(V,E)$ into two sets to maximize the number of edges between the sets.

Initial QAOA implementation on noisy intermediate scale quantum (NISQ) devices were used for some small-scale problems \cite{zhou2020quantum, harrigan2021quantum, earnest2021pulse} due to the challenges of extending it to larger problems related to the depth limitations of the current generation of NISQ devices.
However, with advancements in error mitigation QAOA can be implemented on NISQ devices for problems in some cases involving more than 100 qubits \cite{pelofske2023quantum, sack2024large, barron2023provable, miessen2024benchmarking, montanez2024towards}.
Several variations of QAOA have been developed that attempt to reduce the required hardware resources, such as multi-angle QAOA and XQAOA \cite{herrman2022multi, wurtz2021classically, vijendran2023expressive, shi2022multiangle}, but hardware implementation remains challenging.
As a result, the performance of QAOA is often studied through the classical simulations of the algorithm.
In particular, closed-form equations for the expected value of the MaxCut cost Hamiltonian $C$ have been derived for the special case when only one layer of QAOA is applied \cite{wang2018quantum, herrman2022multi}. Furthermore, more general closed-form equations for all combinatorial optimization problems that have an Ising formulation have also been derived by Ozaeta et. al \cite{Ozaeta_2022}.
Closer inspection of these equations reveals that the expectation value of the cost Hamiltonian may be negatively affected by the presence of triangles in the graph or may be impacted by the maximum degree of the graph.
As such, removing one of the terms that correspond to an edge in a triangle or an edge incident to a maximal degree vertex of the graph might improve the performance of QAOA.
Furthermore, one can consider other modifications of the QAOA circuit, detailed below, and their effect on the performance of QAOA.
One such modification is to use a subgraph of the original graph to generate the circuit.
Using subgraphs for QAOA has been explored in \cite{zhou2023qaoa, li2022large, ponce2023graph} in order to solve large instances of the MaxCut problem using fewer quantum resources.

Generally, the state vector prepared by the QAOA circuit can be written as
\begin{equation*}
    \ket{\vec{\gamma}, \vec{\beta}} = U(B, \beta_p) U(C, \gamma_p) \cdots U(B, \beta_1) U(C, \gamma_1) \ket{s}
\end{equation*}
where $U(B, \beta)$ and $U(C, \gamma)$ are the unitary evolution operators under the action of the corresponding Hamiltonians $B$ and $C$, and $\ket{s}$ is a maximum (or minimum) energy eigenstate of $B$.
The mixer Hamiltonian $B$ is usually defined as a sum of Pauli-X operators on each qubit
\begin{equation*}
    B = \sum X_i,
\end{equation*}
although other mixers have been considered \cite{wang2020x, bartschi2020grover}.
The \textit{phase operator} Hamiltonian $C$ is usually taken to be equal to the \textit{cost} (or phase) Hamiltonian $C'$, i.e. the Hamiltonian whose expectation needs to be minimized by QAOA
\begin{equation*}
    \mathrm{min} \bra{\vec{\gamma}, \vec{\beta}} C' \ket{\vec{\gamma}, \vec{\beta}}.
\end{equation*}
However, the ideas described above make the phase operator different from the cost Hamiltonian ($C \ne C'$), which may potentially improve the performance of QAOA for some choices of the phase operators. 

One similar development has been considered in Ref. \cite{Farhi2017}, where the authors used a Hamiltonian based on hardware connectivity graph as a phase operator, and concluded that a worse, but still non-trivial performance can be achieved with such phase operator. However, they did not try to search for other phase operators that could potentially achieve better performance than the usual cost Hamiltonian.

Other works have tried using different phase operators. In particular, the authors of Ref.~\cite{liu2022quantum} suggested removing the terms of the phase operator corresponding to the edges that are not likely to be in the MaxCut and found that this indeed may improve the performance of QAOA, while also reducing the number of gates in its circuit implementation. Another independent work \cite{wang2023quantum} considered a different set of strategies to choose the terms to remove and arrived at similar conclusions for the case of not-all-equal 3-SAT problems. However, both of these works rely on starting from a classical solution, which results in a poor performance if such solution is far from optimal.
There has also been work on using random gate activation in the setting of Variational Quantum Eigensolver, where in each step of optimization, they activate or unfreeze the trainable parameters of the two-qubit gates \cite{liu2023training}.

In this work, we focus instead on strategies that are based directly on graph structure and do not require classical solutions.
Another important contribution of this work is the analytical formula for the expected value of MaxCut for arbitrary subgraph phase operators in the $p=1$ case, which we derive in Section \ref{sec:randomcircuits}.

In Section \ref{sec:results}, we start by comparing the standard ($C = C'$) formulation to a ``random'' phase operator formulation, where each term in the cost Hamiltonian is selected randomly. Note that one can imagine a phase operator with $Z_iZ_j$ interactions as a graph with edges $Z_iZ_j$, hence we may also talk about phase operator graphs occasionally throughout the text.
We also consider the case where some fraction of the terms in the cost $C'$ Hamiltonian are used to generate the phase operator $C_{\text{sub}}^\alpha$, where $\alpha$ represents the fraction of terms  selected from the original Hamiltonian $C'$.
If $G$ is quite dense and $\alpha$ is small, the number of terms and, correspondingly, gates in the circuit can be significantly reduced, therefore lowering the error in the circuit.
Lastly, there is evidence that triangles in problem graphs affect the QAOA MaxCut approximation ratio \cite{farhi2020quantum, farhi2020quantumwhole, wurtz2021maxcut, herrman2021impact}.
Thus, we consider a special case of the phase operator that corresponds to the original graph $G$ with a subset of triangles removed.
The resulting graph is then used to generate the phase operator.

Finally, we discuss this work and future research directions in Section~\ref{sec:discussion}.

\section{Derivation of analytical MaxCut expected value with custom phase operators}\label{sec:randomcircuits}

For the MaxCut problem, the cost Hamiltonian is defined as:
\begin{equation}
    \label{eq:maxcut_cost}
    C' = \sum_{uv \in E} C'_{uv} = \sum_{uv \in E} \frac{1}{2} (\mathbb{I} - Z_u Z_v)
\end{equation}

Using the Pauli-solver algorithm detailed in \cite{hadfield2018quantum}, the expected value of $\langle C'_{uv} \rangle$ from Eq.~\eqref{eq:maxcut_cost} after 1 layer of QAOA with a random Hamiltonian is:

\begin{equation}
    \bra{s}e^{i\gamma C_{\text{rand}}}e^{i \beta B }C_{uv}e^{-i \beta B }e^{-i\gamma C_{\text{rand}}}\ket{s}
    = \frac{1}{2} - \frac{1}{2}\bra{s}e^{i\gamma C_{\text{rand}}}e^{i \beta B} Z_u Z_v e^{-i \beta B }e^{-i\gamma C_{\text{rand}}}\ket{s}.
\end{equation}

Most terms of the mixer commute to leave

\begin{equation}\label{eq:ev}
    e^{i \beta B} Z_u Z_v e^{-i \beta B } = \cos^2(2\beta) Z_u Z_v+\sin(2\beta)\cos(2\beta)(Y_u Z_v+Z_uY_v)+\sin^2(2\beta)Y_uY_v.
\end{equation}
\noindent
The first term of this sum commutes with $e^{-i\gamma C_{\text{rand}}}$ and does not contribute to the expected value of the sum, i.e.
\begin{equation*}
    \bra{s}e^{i\gamma C_{\text{rand}}} Z_u Z_v e^{-i\gamma C_{\text{rand}}}\ket{s} = 0
\end{equation*}

Let us look at the effect of conjugation on the second term of Eq.~\eqref{eq:ev}

\begin{align*}
 \bra{s}e^{i\gamma C_{\text{rand}}}Y_u Z_ve^{-i\gamma C_{\text{rand}}}\ket{s}.
\end{align*}
$C_{\text{rand}}$ has terms of the form $\frac{1}{2} \left( \mathbb{I} - Z_a Z_b \right)$, however all terms that do not have the form $Z_u Z_c$ for some $c$ commute and cancel, so

\begin{align*}
 \bra{s}e^{i\gamma C_{\text{rand}}}Y_u Z_ve^{-i\gamma C_{\text{rand}}}\ket{s}
 = \bra{s}e^{-i\gamma \sum_{c}(Z_u Z_c)}Y_u Z_v\ket{s}.
\end{align*}

If $Z_u Z_v$ is not a term of $C_{\text{rand}}$, then there is no contribution to the expected value and
\begin{equation*}
    \bra{s}e^{-i\gamma \sum_{c}(Z_u Z_c)}Y_u Z_v\ket{s} = 0
\end{equation*}
Otherwise,
\begin{align*}
\bra{s}e^{-i\gamma \sum_{c}(Z_u Z_c)}Y_u Z_v\ket{s}
   &= \bra{s}e^{-i\gamma Z_u Z_v}e^{-i\gamma \sum_{c \neq v}(Z_u Z_c)}Y_u Z_v\ket{s}\\
   &= \bra{s}(\cos(\gamma) \mathbb{I}-i\sin(\gamma)Z_u Z_v)\prod_{i=1}^d(\cos(\gamma) \mathbb{I}-i\sin(\gamma)Z_u Z_{c_i})Y_u Z_v\ket{s}
\end{align*}
where $d$ is the number of terms of the form $Z_u Z_{c}$ in $C_{\text{rand}}$, with $c \neq v$.
The only term that contributes to the expected value is $-i\sin(\gamma)Z_u Z_v \cos^d(\gamma) Y_u Z_v = -\sin(\gamma)\cos^d(\gamma)X_u$, which implies
\[
   \bra{s} e^{i \gamma C_{\text{rand}}} Y_u Z_v e^{-i \gamma C_{\text{rand}}} \ket{s} = \bra{s}-\sin(\gamma)\cos^d(\gamma)X_u\ket{s} = -\sin(\gamma)\cos^d(\gamma).
\]

By symmetry, the third term of Eq.~\eqref{eq:ev} is also 0 if $Z_u Z_v$ is not a term of $C_{\text{rand}}$, otherwise:
\[
   \bra{s} e^{i \gamma C_{\text{rand}}} Z_u Y_v e^{-i \gamma C_{\text{rand}}} \ket{s} = \bra{s}-\sin(\gamma)\cos^e(\gamma) X_v\ket{s} = -\sin(\gamma)\cos^e(\gamma).
\]
where $e$ is the number of terms of the form $Z_h Z_v$ in $C_{\text{rand}}$, where $h \neq u$.

Hence, we have
\[
    \bra{s} e^{i \gamma C_{\text{rand}}} \left( Y_u Z_v + Z_u Y_v \right) e^{-i \gamma C_{\text{rand}}} \ket{s} = -\chi_{uv} \sin(\gamma) (\cos^d(\gamma) + \cos^e(\gamma)),
\]
where $\chi_{uv} = 1$ if $Z_u Z_v$ is a term in $C_{\text{rand}}$ and 0 otherwise.

Let us look at the effect of conjugation on the last term of Eq.~\eqref{eq:ev}

\begin{align*}
\bra{s} e^{i \gamma C_{\text{rand}}} Y_u Y_v e^{-i\gamma C_{\text{rand}}}\ket{s}
 &=\bra{s} e^{i\gamma \sum_{b \ne v}Z_u Z_b} e^{i\gamma \sum_{a \ne u} Z_a Z_v} Y_u Y_v \ket{s}\\
 &= \bra{s} \prod_{i=1}^{d} (\cos(\gamma) \mathbb{I} - i\sin(\gamma) Z_u Z_{b_i}) \prod_{j=1}^{e} (\cos(\gamma) \mathbb{I} - i\sin(\gamma) Z_{a_j} Z_v) Y_uY_v \ket{s}.
\end{align*}

By approaches similar to those in \cite{hadfield2018quantum}, the expected value contribution from this term is $\frac{1}{2} \cos^{d+e-2f}(\gamma)(1-\cos^f(2 \gamma))$, where $f$ is the number of pairs of terms that satisfy the triangle condition for edge $(u, v)$. We say two terms $Z_u Z_w$ and $Z_w Z_v$ for $w \neq u, v$ in the phase operator $C_{\text{rand}}$ satisfy the \textit{triangle condition} if $(u,v)$ is an edge in the graph of interest.

Combining the above terms yields 
\begin{equation}\label{eqn:ev_final}
    \langle C'_{uv} \rangle = \frac{1}{2} + \frac{\chi_{uv}}{4}\sin(4\beta)\sin(\gamma)(\cos^d(\gamma)+\cos^e(\gamma))-\frac{1}{4}\sin^2(\beta) \cos^{d+e - 2f}(\gamma)(1-\cos^f(2\gamma)).
\end{equation}

\subsection*{Example: An eight-vertex graph where a subgraph phase operator achieves an approximation ratio of 1}

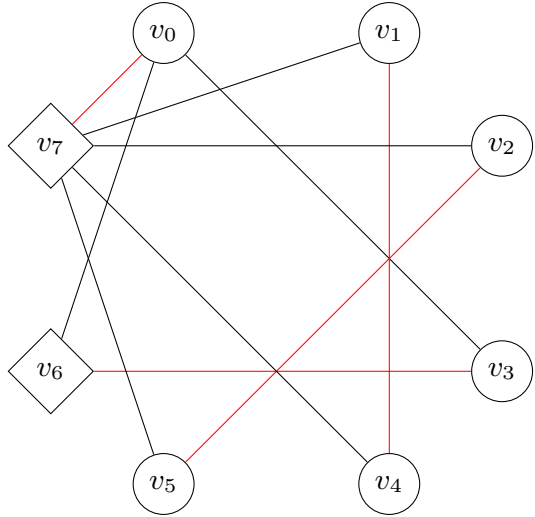
\begin{figure}
 \centering
 \begin{tikzpicture}[scale=1.5]
\begin{scope}[every node/.style={scale=1.25,circle,draw}]
    \node (A) at (0,0) {$v_0$};
    \node (B) at (2,0) {$v_1$};
	\node (C) at (3,-1) {$v_2$}; 
	\node (D) at (3,-3) {$v_3$}; 
 	\node (E) at (2,-4) {$v_4$}; 
  	\node (F) at (0,-4) {$v_5$}; 
\end{scope}

\begin{scope}[every node/.style={scale=1.25,diamond,draw}]
   	\node (G) at (-1,-3) {$v_6$}; 
    	\node (H) at (-1,-1) {$v_7$}; 
\end{scope}

\draw  (A) -- (D);
\draw  (A) -- (G);
\draw[red]  (A) -- (H);

\draw[red]   (B) -- (E);
\draw  (B) -- (H);

\draw[red]   (C) -- (F);
\draw  (C) -- (H);

\draw[red]   (D) -- (G);

\draw  (E) -- (H);

\draw  (F) -- (H);

\end{tikzpicture}
\caption{An eight-vertex graph where a subgraph phase operator achieves an approximation ratio of 1. The subset of edges included in the subgraph phase operator is marked by red color. The diamond and round vertices mark the two subsets that achieve the maximum cut. Note that the red edges form a perfect matching.}
\label{fig:8vertexexample}
\end{figure}

Fig.~\ref{fig:8vertexexample} shows a graph $G$ where there exists a subgraph $G'$ consisting of only the red edges such that using the corresponding phase operator yields a higher approximation ratio compared to the usual QAOA phase operator.
The approximation ratio for this graph using the original QAOA phase operator is 0.934, as found with the code from \cite{maqaoarepo}.

The phase operator in this case is given by:

\begin{equation*}
    C = Z_0Z_7 + Z_1Z_4 + Z_2Z_5 + Z_3Z_6.
\end{equation*}

From Eq.~\eqref{eqn:ev_final}, one can see that the cost expectation of edges in $G$ that are not in $G'$ is 0.5. 
The edges that are in $G'$ do not form triangles, therefore it is easy to see that their cost expectation becomes equal to 1 for $\gamma = \pi/2$ and $\beta = \pi/8$. There are 6 edges that are in $G$ but not in $G'$ and 4 edges in $G'$, therefore the total cost expectation, 7, in this case coincides with the maximum cut for this graph.

\subsection*{Example: A family of graphs and circuits with an approximation ratio of 0.5}\label{subsec:starexample}

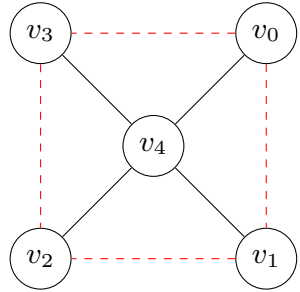
\begin{figure}
  \centering
  \begin{tikzpicture}[scale=1.5]
 \begin{scope}[every node/.style={scale=1.25,circle,draw}]
     \node (A) at (0,0) {$v_4$};
     \node (B) at (1,1) {$v_0$};
 	\node (C) at (-1,1) {$v_3$}; 
 	\node (D) at (-1,-1) {$v_2$}; 
  \node (E) at (1,-1) {$v_1$}; 
 \end{scope}

 \draw  (A) -- (B);
 \draw  (A) -- (C);
 \draw  (A) -- (D);
  \draw  (A) -- (E);
 
  \draw[dashed, red](B) -- (E);
 \draw[dashed, red]  (B) -- (C);
 \draw[dashed, red]  (C) -- (D);
  \draw[dashed, red]  (E) -- (D);

 \end{tikzpicture}
 \caption{The star graph, which consists of vertices and the solid edges. The dashed red edges are the terms included in the phase operator. This choice of phase operator yields an approximation ratio of 0.5.}
 \label{fig:star}
\end{figure}

The star graph on $n$ vertices is a connected $n$-vertex graph that has exactly one vertex of degree $n-1$ and $n-1$ vertices of degree one. 
An example of a star graph is depicted in Fig.~\ref{fig:star}.

Consider a star graph with 3 or more leaves.
Label the center vertex of the star $n-1$ and the other vertices $0$ to $n-2$ in clockwise order, starting at some arbitrary leaf.
Now consider the phase operator with the terms of the form $Z_i Z_{i+1}$ modulo $n-2$, which has $n-1$ terms, the same as the cost Hamiltonian.
With this choice of the phase operator, the last term of Eq.~\eqref{eqn:ev_final} for all edges of the graph is 0, since none of them satisfy the triangle condition.
Furthermore, the second term in Eq.~\eqref{eqn:ev_final} is also zero since none of edges are chosen as a term in the phase operator.
Therefore, all edges in the graph have the cost expectation of 0.5, and the overall cost expectation for a star graph with this phase operator is $\frac{|E|}{2}$.
In general, if the edges of the graph do not correspond to the terms of the phase operator and do not satisfy the triangle condition, the approximation ratio will always be 0.5.
Note if the star has two leaves, this phase operator construction contains a multi-edge, and thus requires additional parameter optimization considerations.

\section{Results}\label{sec:results}
In this work, we considered QAOA with random phase operators, subgraph phase operators, a special case of subgraph phase operators in which a subset of triangles are removed (abbreviated TR-phase operator for triangle-removed), and a subcase of random subgraph where edges that are incident to maximal degree vertices are removed (abbreviated MDER-phase operator for maximum degree edge removal) for solving the MaxCut problem on all 11,117 non-isomorphic eight vertex graphs.
The simulations at $p=1$ were performed by maximizing the expected value formula in Eq.~\eqref{eqn:ev_final} over the parameters $\gamma$ and $\beta$.
The $p=2$ and $p=3$ code and classical optimization subroutine can be found at \cite{wilkiecode}.
The optimizer used was the L-BFGS algorithm with 100 random parameter initialization seeds.

\subsection{Random phase operators}
First, we consider random phase operators and analyze the effect of phase operator structure on the approximation ratio. 
In order to create the phase operator graphs, we randomly selected $m$ of the $8 \choose 2$ possible edges of the complete graph on eight vertices, where $m$ is the number of edges in the original graph.
Mathematically, this is given by
\begin{equation*}
    C_{\text{rand}} = \sum_{ab} \frac{1}{2} \left( \mathbb{I} - Z_a Z_b \right),
\end{equation*}
where the indices $(a, b)$ are random.
For each graph, up to 10 non-isomorphic random phase operators were considered. 
Note that in some cases, such as the complete graph $K_8$, we cannot select 10 non-isomorphic phase operators, since we keep the same number of edges in the phase operator graph. 
However, when possible, we generated 10 non-isomorphic random phase operators and evaluated the approximation ratio for each.

\begin{figure}
    \centering
    \includegraphics[width=0.99\linewidth]{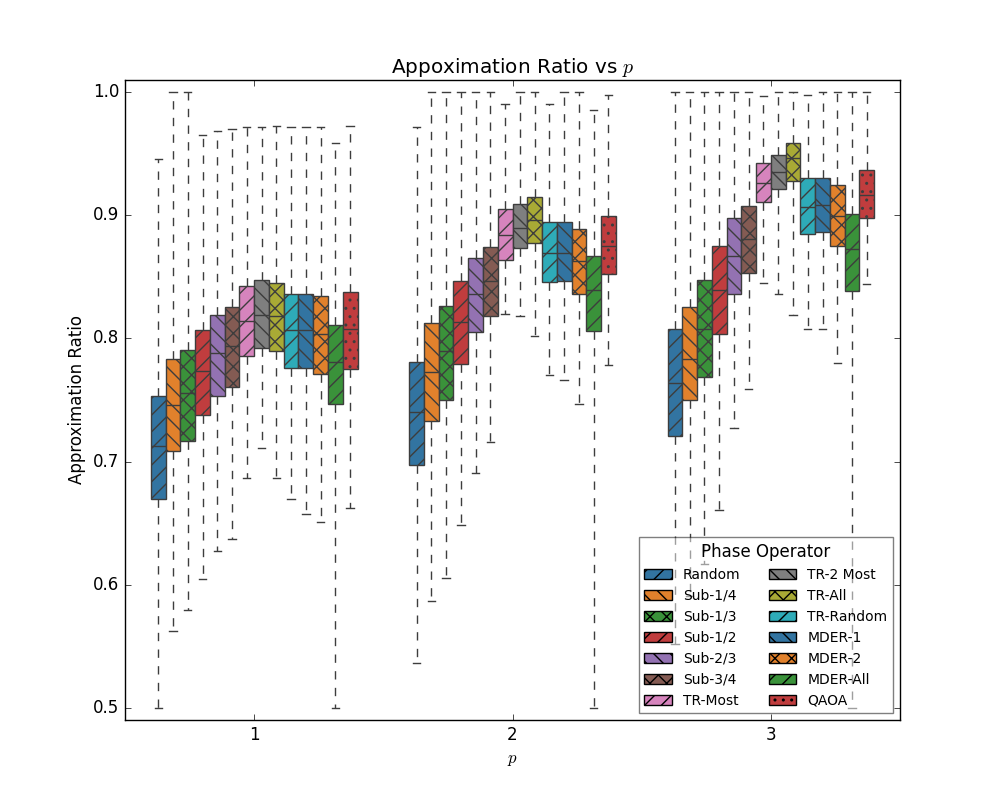}
    \caption{Comparison of the AR for each phase operator and $p$ tested. For each $p$, the Random phase operator performs the worst, while one of the TR-phase operators performs the best, with regular QAOA in the middle.}
    \label{fig:ar_boxplot}
\end{figure}

\begin{figure}
    \centering
    \includegraphics[width=0.99\linewidth]{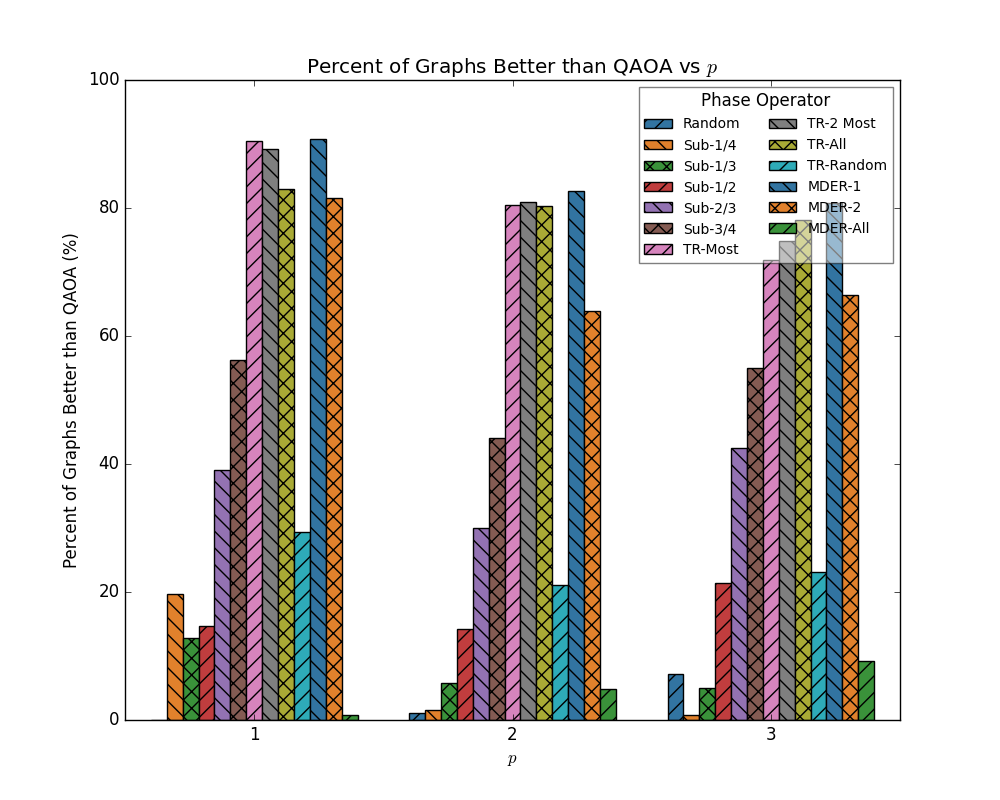}
    \caption{The percentage of 8-vertex graphs that had at least one instance of the phase operator perform better than regular QAOA in terms of approximation ratio.}
    \label{fig:percent_better}
\end{figure}

From Fig.~\ref{fig:percent_better}, we can see that in almost no case did a random phase operator outperform the cost phase operator.
Specifically 0.036\% of the graphs tested at $p = 1$ had a random phase operator that outperformed the cost phase operator.
The random graphs are not guaranteed to have any structure with respect to the original graph, so it is not surprising that the random phase operators typically perform worse.
Furthermore, there are far more than 10 non-isomorphic phase operators that can be generated in most cases, so it is of course possible that the remaining graphs have better phase operators.
Fig.~\ref{fig:ar_boxplot} shows the performance of the random phase operators as compared to the cost phase operator for each $p$.
On average, the random phase operators not only performed worse than the cost phase operator, but also worse than all the other phase operators considered in this work, with the lowest average approximation ratio.
This trend continued for all $p$ tested.

\subsection{Subgraph phase operators}
The subgraph phase operators differ from the purely random ones in that we choose a random subgraph of the original graph $G$ and add to the phase operator $Z_iZ_j$ only the terms corresponding to edges of this subgraph.
In other words, only a subset of terms from the cost Hamiltonian can be considered for the phase operator.

In this experiment, we tested 5 cases of the subgraph phase operator, where the subgraphs chosen contain $\lceil \alpha |E(G)| \rceil$ edges of $G$ for one $\alpha \in \left\{ \frac{1}{4}, \frac{1}{3}, \frac{1}{2}, \frac{2}{3}, \frac{3}{4} \right\}$.
We label each case as Sub-$\alpha$, for the corresponding $\alpha$.
Similar to before, we choose up to 10 non-isomorphic subgraphs for each $\alpha$ and use 100 parameter initialization seeds.

Fig~\ref{fig:ar_boxplot} shows the performance of the subgraph phase operators as compared to the cost Hamiltonian for MaxCut for each $p$.
We can see that the subgraph phase operators performed better than the random phase operators, but not as well as the TR-phase or the MDER-phase operators.
What is more interesting to note is that all the subgraph phase operators performed worse than normal QAOA for every $p$ tested.
We can see from Fig.~\ref{fig:percent_better} that for Sub-$\frac{1}{4}$ and Sub-$\frac{1}{3}$ the percentage of graphs that had at least one phase operator that performed better than normal QAOA decreases with $p$, while for the other cases there is a decrease at $p=2$ and an increase at $p=3$.
Out of all the subgraph phase operators tested, 75.9\% of the considered graphs had at least one subgraph phase operator with a higher approximation ratio at $p=1$.
These results suggest that randomly choosing a subgraph to create the phase operator from does not always yield a better approximation ratio.
Since only 10 non-isomorphic subgraph phase operators were tested, it is possible that there exist phase operators that have even better performance.
Specifically, as we will see with the TR-phase operators in Sec.~\ref{subsec:tr-phase}, subgraphs that have specific edges removed to reduce the number of triangles can improve the approximation ratio.

One interesting note is that there were 10 phase operator subgraphs that achieved an AR of 1, meaning that the optimal MaxCut value was found.
The structure of these subgraphs all have the same form, where the subgraph is a perfect matching as in Fig.~\ref{fig:8vertexexample}. 
A perfect matching $M \subset E$ is a subset of graph edges such that every vertex in $G$ is incident to exactly one edge in $M$. While interesting to note, this condition is not necessary or sufficient to define subgraph phase operators that achieve an approximation ratio of 1. 
However, if the MaxCut value is $\frac{|V|}{4} + \frac{|E|}{2}$, then a perfect matching phase operator, if it exists, does yield an approximation ratio of 1.
In general, this information is not known a priori.


\subsection{TR-phase operators}\label{subsec:tr-phase}
The TR-phase operators are a special case of the subgraph phase operators in which one or more triangles is removed from the original graph.
All edges that are not in triangles remain in the phase operator.
The importance of this case is to reduce the number of triangles so that the last term of Eq.~\eqref{eqn:ev_final} is $0$ for a larger number of edges in the graph, with the hope is that this increases the overall expected value.
This experiment examines four types of TR-phase operators.
The first type, named ``TR-Most'', removes a single edge that is contained in the most triangles in the graph.
The second type, named ``TR-2 Most'', implements the same ``most'' rule, but applied consecutively 2 times.
The third type, named ``TR-All'', iteratively removes edges until there are no more triangles left in the graph.
The final type, named ``TR-Random'', just chooses a single random edge included in a random triangle and removes it.

Fig~\ref{fig:ar_boxplot} shows the performance of the TR-phase operators as compared to the cost phase operators for each $p$.
These results indicate that removing triangles from the cost phase operator can improve the approximation ratio, while also using fewer gates in the circuit.
The TR-phase operators (excluding TR-Random) performed the best over all compared to the other phase operators, which becomes more apparent as $p$ increases.
Out of all the TR-phase operators tested at $p=1$, 95.1\% of the considered graphs had at least one TR-phase operator that yielded a higher approximation ratio.
From Fig~\ref{fig:percent_better}, we see the same trend that TR-All begins to perform better on more graphs than TR-Most and TR-2 Most as $p$ increases.
However, comparing the performance of TR-Most, TR-2 Most, and TR-All, we see that removing all triangles did not yield the best performance initially, but at $p > 2$ TR-All outperforms the other two.
This may be due to the fact that removing all triangles may result in new structure, such as an increase in the diameter of the graph, that may be difficult for QAOA to see.
Additionally, the expectation of the removed edges tends to decrease, which offsets the gain from the remaining edges.
We can also see that for each $p$, removing at least 2 edges performs better on average than removing only a single edge.
In every case, the TR-Random phase operator performed the worst among the TR-phase operators, but still comparable to regular QAOA, which implies that edges should be strategically removed.

\subsection{MDER-phase operators}
The MDER-phase operators are subcases of the subgraph phase operator in which at least one edge that is incident to maximum degree vertices are removed from the original graph.
We examine this type of phase operator because the exponents of the trigonometric functions in the expected value equation depend on degree of the vertex incident to each edge.
Removing edge with high degree drives down the degree of a subset of the trigonometric functions in Eq.~\eqref{eqn:ev_final}.
This may slightly increase the expected value of all edges that share a common vertex with the removed edge at the cost of setting the expected value of the removed edge to 0.5 plus possibly the triangle term in the expected value.
Thus, edges incident to high degree vertices are removed.
This experiment considers 3 types of MDER-phase operators: MDER-1-phase operator where we find the maximal degree vertex and remove a random edge incident to it, MDER-2-phase operator, which consecutively applies the previous strategy 2 times, and MDER-all-phase operator where we find the maximal degree vertex and remove all the edges incident to it.

As shown in Fig~\ref{fig:percent_better}, over $60\%$ of the tested graphs had at least one MDER-1 and MDER-2 instance that gave a higher approximation ratio than QAOA, while almost no MDER-All phase operators did.
Furthermore, MDER-1 and MDER-2 result in approximation ratios comparable to those of QAOA, while MDER-All performed worse at $p=1$,  as seen in Fig~\ref{fig:ar_boxplot}.
However, as $p$ increases the MDER-phase operators begin to perform increasingly worse than regular QAOA.
These results indicate that removing edges incident to maximum degree vertices can yield comparable performance at lower depth to QAOA while using fewer gates.
However, removing all the edges incident to the maximum degree vertex can lead to worse performance.




\section{Discussion}\label{sec:discussion}
In this work, we examined the impact of QAOA phase operator design on the approximation ratio on all non-isomorphic, connected eight-vertex graphs and up to three layers of QAOA.
In general, removing a subset of triangles from the phase operator tends to yield a higher average approximation ratio, and using a random phase operator tends to result in a lower approximation ratio than the approximation ratio yielded by the cost phase operator.

The main implication of this work is that, on average, fewer gates can be used in the QAOA circuit and a comparable approximation ratio can be attained. 
This can provide a modest noise reduction when implementing the algorithm on NISQ devices.
Future work may include multiple directions.
First, one can examine how the subgraph phase operators affect the optimization landscape, as there could be a risk that modifying the phase operator results in more barren plateaus.
Second, it is important to consider the scalability of the custom phase operators with respect to the number of layers of QAOA.
Another direction includes characterizing of when removing gates will improve the approximation ratio.
We will also consider using MA-QAOA with the subgraph phase operators to see if the approximation ratio of a given problem can be increased even further.
Since the subgraph phase operators have fewer gates, there would be fewer angles to optimize. Thus, the parameter optimization of these multi-angle circuits should be easier, which overcomes one of the challenges of MA-QAOA implementation \cite{gaidai2023performance}.
It would also be interesting to choose other types of phase operators.
In particular, at iteration $p$ of QAOA, cycles of length $2p+1$ contribute to the expected value of edges in the cycle.
One could study how removing edges from appropriate length cycles at each iteration of QAOA impacts the approximation ratio.
We believe this type of dropout method would result in even higher approximation ratios than just removing edges in triangles, as in this work.

One possible application of this work is in the area of privacy.
In \cite{upadhyay2023obfuscating}, the authors explore creating multiple circuits with various gates removed from the QAOA cost Hamiltonian in order to protect the privacy of the problem at hand at the cost of having to run multiple circuits on different third party hardware and recombining the results.
Our work shows that we can remove certain gates from the cost Hamiltonian, thus increasing the privacy of the original problem, and have comparable results without relying on multiple circuits.

\section*{Acknowledgements}
J.O. acknowledges DARPA ONISQ program under award W911NF-20-2-0051. R. H., J. O., and A. W. acknowledge NSF CCF 2210063.

\renewcommand\refname{References Cited}
\bibliography{references}
\bibliographystyle{IEEEtran}

\end{document}